# Dual-Energy Subtraction Efficiency: Development of an Objective Quality Metric for Tissue-Subtraction Radiography

Sebastian Lopez Maurino [a)]
*Department of Physics, KA Imaging Inc., 3-560 Parkside Dr, Waterloo, Ontario N2L 5Z4, Canada*

Karim S. Karim
*Department of Electrical and Computer Engineering, University of Waterloo, 200 University Ave W, Waterloo, Ontario N2L 3G1, Canada*

**Purpose:** We propose a new objective numerical figure of merit to aid in the evaluation and comparison of tissue-selective images generated from dual-energy radiography systems.

**Methods:** A metric is developed through identification of the requirements of a successful objective metric and analyzed in a variety of scenarios

**Results:** The proposed Dual-Energy Subtraction Efficiency (DSE) metric adequately describes a multitude of image properties for all simulated image scenarios, indicating its ability to accurately and objectively describe image quality

**Conclusions:** The DSE and its measurement method described can become a tool to characterize dual-energy radiographic image quality objectively and quantitatively, allowing for improved system comparison, development, and optimization

## I. INTRODUCTION

Dual-Energy (DE) radiography has found a variety of application in the medical imaging space with well-known clinical benefits[1–9]. The recent widespread adoption of digital X-ray detectors, together with advances in X-ray generators and a reduction in component cost, has resulted in an increased number of DE offerings by a variety of equipment manufacturers[10–12]. With this, there follows a clinical, industrial, and regulatory need for an objective method of comparing image quality across current and future offerings. However, this has proven to be challenging due to the wide variety of technological approaches to DE imaging –from the method of DE data acquisition[10,13–15] to the tissue-subtraction algorithms used[16–20] to noise reduction techniques[21–23]–, as well as the complexity of the tissue-subtracted images obtained through the DE process[24,25]. As a result, many methods have been used in literature to evaluate and compare DE image quality[26–30], each with their own set of deficiencies.

Our objective is to develop an experimental and analytical method to objectively quantify the clinically-relevant properties of tissue-subtracted DE images. The resulting metric can then be used to compare current technologies, can serve as a regulatory and quality assurance tool, and can aid in the development and optimization of future technologies.

## II. METRIC REQUIREMENTS

To motivate the specifics of the quality metric measurement and computation process, a series of requirements were imposed that would result in a successfully objective and complete metric. These were inspired by the common clinical applications of DE radiography, by the variety of existing technologies, and by the important objectivity property. The developed quality metric must

**be independent of imaging and subtraction method** – of course, the developed metric should be able to be applied to a variety of both DE image data acquisition methods and tissue-subtraction algorithms. Furthermore, it should serve as an effective tool to compare such technologies, and accurately describe their respective benefits and drawbacks;

**reflect the amount of tissue subtraction** – the main purpose of DE tissue subtraction radiography is to reduce or remove anatomical noise, thereby enhancing visibility of the anatomies of interest. The quintessential example of this process is the removal of the overlying ribs in a chest radiograph in order to more clearly focus on the lung field. The developed metric should quantify the quality of this anatomical noise removal, and drop from its optimal value when residual noise is still present in the tissue subtracted image;

**quantify the image noise properties and resolution** – many design decisions in radiography (such as pixel size, scintillator thickness, input X-ray spectrum properties, etc.) encompass a compromise between image noise and resolution. This is intensified in tissue subtraction DE imaging due to the image-processing nature of this technique. It is important that the developed metric closely quantifies both noise and resolution in the tissue-subtracted image, since a metric that only reflects one of these could easily be optimized at the detriment of the other;

**be possible to measure in a standardized, reproduceable way** – the experimental setup must be simple enough such that any modestly-equipped laboratory may be able to reproduce it, and the data analysis strictly standardized so as to not leave room for uncertainty caused by mathematical implementations. Moreover, the process must be designed to accommodate any commonly-used data type and still provide universally comparable results;

**be immune to a variety of edge cases** – the complex nature of tissue subtraction techniques can result in images that may, at their surface, look to have improved quality, but in reality have done so at the expense of added artifacts to the image. A common example of such artifacts are edge artifacts introduced at the boundary between two tissue types, typically caused by a mismatch in the effective resolution of the input DE image data. A successful metric should recognize and quantify such image artifacts;

## III. PROPOSED METRIC: DUAL-ENERGY SUBTRACTION EFFICIENCY

We propose a new metric, coined Dual-Energy Subtraction Efficiency (or $DSE$), which through its experimental procedure and image analysis, can achieve the stated goals. It is based on the spectral transfer of noise, contrast, and resolution properties from the captured (or primary) DE image data to the generated (or secondary) tissue-subtracted images.

### III.A. Experimental Setup

The process of measuring $DSE$ begins by obtaining the primary image data. These are the images obtained directly through X-ray exposures. Depending on the Dual-Energy imaging system used, an acquisition of a Dual-Energy data set may consist of one or many X-ray exposures. Typically, spectral switching systems will perform two or more X-ray exposures separated in time and using differing in spectral properties, such as X-ray source kilovolt peak and/or additional beam filtration. On the other hand, single-exposure systems obtain a multitude of images of different spectral properties through energy-resolving capabilities of their X-ray detector, the most common of which are multi-layer devices, which consist of two or more stacked X-ray sensors, sometimes separated by a beam-hardening filter.

The required data sets for the calculation of the $DSE$ are images of a step phantom. This phantom consists of a 50 mm acrylic (polymethyl methacrylate) base and two interchangeable edges: an acrylic edge and an aluminum edge. The base is meant to simulate a constant amount of background soft-tissue similar to that of a lung field, while the edges are used to introduce soft-tissue and hard-tissue contrast, respectively. Acrylic and aluminum were selected as the phantom materials due to their similar X-ray attenuation properties to soft-tissue and bone respectively, and due to their universal availability. The phantom is further parametrized by the thickness of the edges, $t$: the acrylic edge is defined to have a thickness of $5t$ mm, while the aluminum is defined to have a thickness of $t$ mm. This $5:1$ ratio was selected given the approximate $5\times$ expected attenuation of aluminum versus acrylic[1].

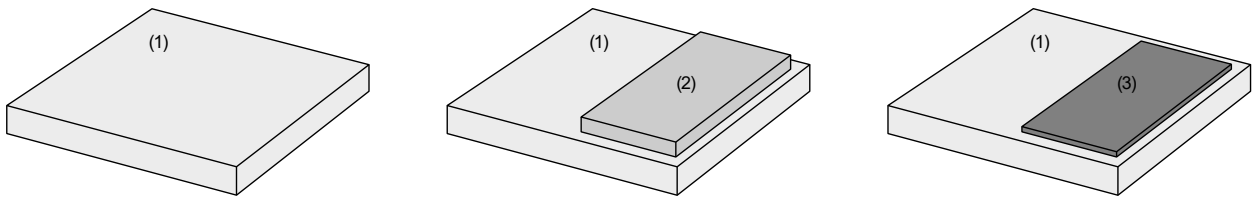

Fig. 1 DSE phantom consisting of an acrylic base (1), an acrylic edge (2) and an aluminum edge (3). Shown in its base-only configuration (left), its acrylic-edge configuration (centre), and its aluminum-edge configuration (right).

Six separate primary data sets are required for the measurement of a system's $DSE$ in one dimension. These are denoted with $\mathbb{I}_{p,k}$, where the subscript $p \in \{b, s, h\}$ indicates the phantom configuration being imaged (base only, base and acrylic edge, base and aluminum edge, respectively), and the subscript $k \in \{X_1, X_2\}$ indicates the exposure level used to acquire the image, where $X_1$ refers to air kerma levels commonly used by this system in a clinical setting, and $X_2$ refers to approximately double that. Note that the symbol $\mathbb{I}$ is chosen to indicate that a data set contains two or more images which differ from each other in their X-ray spectral (i.e. energy) properties.

---

[1] for TASMICS spectra with 1.6 mm intrinsic aluminum filtration and a tube kilovolt peak ranging from 40 kV to 150 kV, the average beam, after being attenuated by a 50 mm acrylic base, will see the same energy attenuation from 1 mm of aluminum as from 6.55 mm of acrylic. Note that this value varies significantly throughout this spectral range, and the $5:1$ ratio is meant solely as a simple first-order approximation of relative attenuations, not as a precise value.

In order to appropriately measure a supersampled edge profile of the phantom, careful placement of the phantom in the X-ray system geometry is required. The phantom must be placed immediately on the surface of the X-ray detector, with the phantom edge creating an angle with the detector active array of 1° to 3°, and the X-ray beam line as close as possible to the edge centre.

Table 1 List of primary data sets obtained directly through X-ray exposures, each with differing phantom configuration and air kerma level.

| Air Kerma Level | Phantom configuration | | |
| --- | --- | --- | --- |
| | Base only | Acrylic edge | Aluminum edge |
| ~ Typical Clinical Setting | $\mathbb{I}_{b,X_1}$ | $\mathbb{I}_{s,X_1}$ | $\mathbb{I}_{h,X_1}$ |
| ~2 × Typical Clinical Setting | $\mathbb{I}_{b,X_2}$ | $\mathbb{I}_{h,X_2}$ | $\mathbb{I}_{h,X_2}$ |

### III.B. Derived Images and Dual-Energy Subtraction

Once the primary data sets have been obtained, it is possible to mathematically combine them to obtain derived images. Two types of derived images are of interest to $DSE$ calculation: summation images, and tissue-subtracted images. Summation images are a simple addition of all images in a primary data set, and represent the total signal obtained in all exposures pertaining to a single data set. Tissue-subtracted images are those obtained through dual-energy processing of the primary data set with the goal removing all contrast from a particular tissue type or material. The most common approach for obtaining subtracted images is through logarithmic subtraction, but any method of contrast cancellation is acceptable for the calculation of $DSE$, as long as it is an algorithmic combination of the images in the primary data set.

Three derived images need be obtained from each primary data set using these methods. These are denoted with $I'_{d,p,k}$, where the prime symbol is used to indicate their computed nature, and the subscript $d \in \{\Sigma, H, S\}$ indicates the image is a summation, hard-tissue only (i.e. soft-tissue subtracted), or soft-tissue only (i.e. hard-tissue subtracted) image, respectively. Table 2 below shows a summary of all the necessary derived images. Note that when obtaining the derived images, the exact same processing must be applied to images corresponding to different phantom configurations and exposure levels.

Table 2 List of derived images obtained through summation and dual-energy subtraction from the primary data sets.

| Derived image type | Phantom configuration | | |
| --- | --- | --- | --- |
| | Base only | Acrylic edge | Aluminum edge |
| Summation | $I'_{\Sigma,b,X_1}$, $I'_{\Sigma,b,X_2}$ | $I'_{\Sigma,s,X_1}$, $I'_{\Sigma,s,X_2}$ | $I'_{\Sigma,h,X_1}$, $I'_{\Sigma,h,X_2}$ |
| Soft-tissue subtraction (i.e. hard-tissue image) | $I'_{H,b,X_1}$, $I'_{H,b,X_2}$ | $I'_{H,s,X_1}$, $I'_{H,s,X_2}$ | $I'_{H,h,X_1}$, $I'_{H,h,X_2}$ |
| Hard-tissue subtraction (i.e. soft-tissue image) | $I'_{S,b,X_1}$, $I'_{S,b,X_2}$ | $I'_{S,s,X_1}$, $I'_{S,s,X_2}$ | $I'_{S,h,X_1}$, $I'_{S,h,X_2}$ |

Due to the loss of the linear relationship between input air kerma and image values caused by most tissue subtraction algorithms, it is necessary to renormalize all tissue-specific derived images. This is accomplished by linearly rescaling the image such that the median signal values in the base and tissue-specific step regions match those of its corresponding summation image[2]. For example, $I'_{H,p,X_1}$ are rescaled using

$$I''_{H,p,X_n} = \left(I'_{H,p,X_n} - \tilde{\mu}_{H,b,X_n}(b)\right) \times \frac{\tilde{\mu}_{\Sigma,h,X_n}(s) - \tilde{\mu}_{\Sigma,h,X_n}(b)}{\tilde{\mu}_{H,h,X_n}(s) - \tilde{\mu}_{H,h,X_n}(b)} + \tilde{\mu}_{\Sigma,b,X_n}(b)$$

where $I''_{d,p,X_n}$ is the obtained renormalized derived tissue-specific image, and $\tilde{\mu}_{d,p,X_n}(b,s)$ is the median value of the base or step region (respectively) in the derived image of the same subscript. The motivation behind this renormalization technique is that, in dual-energy images, the magnitude of the image noise only makes sense in relation to the contrast available in the remaining tissue type. Unlike conventional radiography images, dual-energy images do nor contain a reference point of zero signal, and thus it is necessary to utilize two arbitrary signal points to define a relevant contrast. The remaining phantom step is the natural choice for such signal points.

---

[2] This renormalization procedure presumes that the summation images follow a linear relationship between their image values and the input air kerma. If this is not found to be the case, a linearization function should first be applied to all summation images. This function should be computed by measuring the average signal for summation images containing the phantom in its base-only configuration.

### III.C. Definition

The Dual-Energy Subtraction Efficiency metric was developed such that it can characterize three aspects of the tissue-subtracted images:

- the relative noise properties of the tissue-subtracted images with respect to the primary data set;
- the relative spatial resolution of the tissue-subtracted images with respect to the primary data set;
- the quality of tissue cancellation.

In this way, $DSE$ reflects the transfer of image quality metric from the primary data set to the tissue-subtracted images, and not necessarily the absolute quality of the image. This metric may be used in conjunction with a more objective metric such as modulation transfer function and detective quantum efficiency to obtain a full picture of system resolution and dose efficiency.

Mathematically, the $DSE$ is defined in terms of the noise transfer $NT_d(u,v)$, the resolution transfer $RT_d$, and the subtraction artifact power $AP_d$ as

$$DSE_d(u,v) = \left(1 - \sqrt{AP_d(u,v)}\right)\frac{RT_d^{\,2}(u,v)}{NT_d(u,v)}$$

where $d \in \{H, S\}$ indicates the tissue-type subtracted. These and their computation procedures are further described in the following sections. In this way, the $DSE$ represents the conjunction of all three image characteristics it aims to describe.

### III.C.1. Noise Transfer

The noise transfer term measures the increase in quantum noise that results from dual-energy subtraction. This is accomplished by measuring the spectral power in a quantum-noise only image in both the tissue-subtracted image and the summation image. The image of the phantom in its base-only configuration was chosen to represent a typical amount of background quantum noise. Therefore, $NT_d$ is defined as

$$NT_d(u,v) = \frac{NPS_{d,b,X_1}(u,v)}{NPS_{\Sigma,b,X_1}(u,v)}$$

where $NPS_{d,p,X_n}(u,v)$ is the noise power spectrum of the derived image of the same subscript, which must be first renormalized in the case of tissue selective images.

### III.C.2. Spatial Resolution Transfer

In order to measure the effects that dual-energy subtraction has on image resolution, we can study the changes in the edge profile created by the phantom step. A subtraction algorithm that results in a loss of resolution will see a spreading of this edge profile. It is therefore possible to quantify the loss in resolution as a function of spectral frequency with

$$RT_d = \frac{SSF_{S,p,X_1}(u,v)}{SSF_{\Sigma,s,X_1}(u,v)}$$

for $p = s$ when $d = S$ and $p = h$ when $d = H$, and where $SSF_{d,p,X_n}(u,v) = |\mathcal{F}(LSF_{d,p,X_n})|$ is the step spectral function and LSF is the line spread function of the edge generated by the phantom step.

### III.C.3. Tissue Cancellation

In order to measure the quality of tissue cancellation, we can analyze the spectral properties of the remaining artifacts after subtraction. However, obtaining a power spectrum of the tissue subtracted images will show contributions from both subtraction artifacts and image quantum noise. In order to decouple them, the tissue-subtracted images are obtained at two separate exposure levels: $X_1$ and $X_2$. In practice, these two exposure levels are selected to achieve a $\sim 2 : 1$ ratio, but any ratio may suffice to accurately measure artifact contributions.

This decoupling is possible thanks to the differing relationships the spectral power of each of component has to input air kerma levels. Due to the poissonic nature of quantum noise, all spectral power components caused by it will increase linearly with input exposure levels, while image artifacts are assumed to be a linear function of input signal, and thus their spectral power components will increase with the square of input exposure level.

It is therefore possible to obtain an estimate of the spectral power contributions of the remaining artifacts using

$$AP_d(u,v) \approx \frac{EPS_{d,p,X_2}(u,v) - (X_2/X_1) \times EPS_{d,p,X_1}(u,v)}{\overline{AP\Sigma_p}(u,v)}$$

$$AP\Sigma_p(u,v) \approx EPS_{\Sigma,p,X_2}(u,v) - (X_2/X_1) \times EPS_{d,p,X_1}(u,v)$$

for $p = h$ when $d = S$ and $p = s$ when $d = H$, and where $EPS_{d,b,X_2}(u,v)$ is the edge power spectrum representing the spatial spectral power of the corresponding image measured around the step edge. In this way, $AP_d(u,v)$ represents the portion of step signal remaining in the dual-energy image, since it is normalized to the power of the same edge in the summation image. Note that the summation image power is averaged across all frequencies, since is it possible that the subtraction process includes a frequency transfer of the step features, and thus the edge artifact components in the dual-energy image may be of higher power than the full step in the summation image for some frequencies.

## IV. RESULTS

To evaluate the proposed metric, mathematical simulations of a two-shot dual-energy system were used to assess a variety of acquisition and subtraction scenarios in order to ensure that the $DSE$ reflects the final image quality. These simulations serve to illustrate the results of this metric when applied to an idealized system, which utilizes a perfectly absorbing detector and lacks all system nonidealities such as scatter, input beam nonuniformities, geometric misalignment, etc.

Table 3 Results of mathematical simulations of an ideal X-ray systems in a variety of dual-energy subtraction scenarios. Simulated system consisted of an ideal X-ray detector of 100 μm square pixels with a scintillator blur simulated using a $\sigma = 0.45$ gaussian blur kernel. Two exposures were simulated at 60 kVp and 120 kVp at an input air kerma of 0.1 mGy and $0.0\hat{6}$ mGy respectively, achieving a 1 : 1.5 ratio in low- to high-energy exposure. The DSE phantom was simulated in its $t = 2$ configuration with a 2° rotation from the detector array axis. All simulations generated a $512 \times 512$ image, which were repeated 32 times and their metrics averaged to obtain the values shown below. Metric graphs show the $DSE_S$ (solid) and its $NT^{-1}$ (dotted) term on the left axis and its $RT^2$ (dot-dashed) and $1 - \sqrt{AP}$ (dashed) terms on the right axis.

| Scenario | Sample $I''_{H,s,X_1}$ | Metrics | Discussion |
|---|---|---|---|
| Ideal | 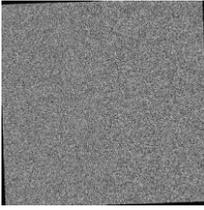 | 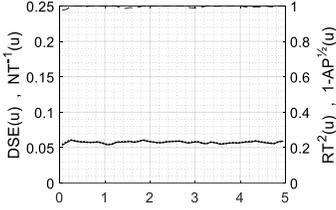 | This scenario illustrates an ideal DE logarithmic subtraction. The $RT$ and $AP$ terms $= 1 \, \forall \, u$ in this ideal situation, and thus the $DSE$ is only affected by $NT$, which indicates that the DE images' $SNR^2$ is ~6% that of the summation image. |
| Imperfect Subtraction | 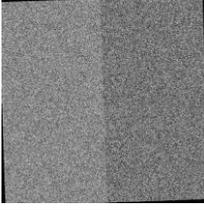 | 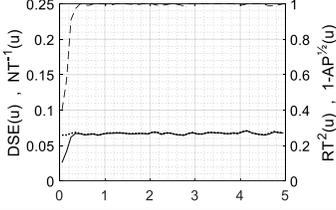 | This scenario shows a nonideal selection of subtraction factor in the DE logarithmic subtraction. While this results in a higher image SNR as shown by the higher $NT$ term, the remaining edge artifact in $I''_{H,s,X_1}$ causes a large decrease of the $AP$ term at low frequencies, decreasing the $DSE$ there. |

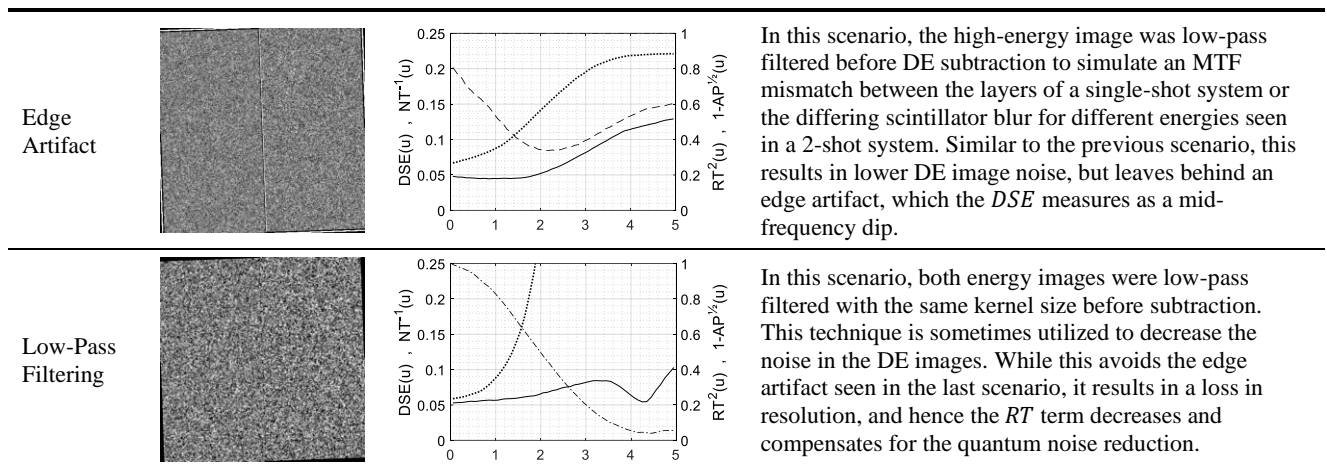

| | | | |
|---|---|---|---|
| Edge Artifact | | | In this scenario, the high-energy image was low-pass filtered before DE subtraction to simulate an MTF mismatch between the layers of a single-shot system or the differing scintillator blur for different energies seen in a 2-shot system. Similar to the previous scenario, this results in lower DE image noise, but leaves behind an edge artifact, which the $DSE$ measures as a mid-frequency dip. |
| Low-Pass Filtering | | | In this scenario, both energy images were low-pass filtered with the same kernel size before subtraction. This technique is sometimes utilized to decrease the noise in the DE images. While this avoids the edge artifact seen in the last scenario, it results in a loss in resolution, and hence the $RT$ term decreases and compensates for the quantum noise reduction. |

## V. DISCUSSION

It is clear from these simulation results that the $DSE$ metric appropriately describes the desired dual-energy image properties, namely the amount of tissue subtraction, the noise and resolution properties, and the presence of subtraction artifacts. It is even possible to identify the particulars of the image deficiency by studying the $DSE$ spectral shape.

However, the complexity caused by the differing spectral characteristics may make it difficult to use the $DSE$ as a tool when doing large optimizations of dual-energy systems, particularly given the dependency of this metric on the phantom configuration. To address this, we propose the use of the minimum $DSE$, or $mDSE$ (defined as the minimum value of the $DSE$ across the entire frequency spectrum), to serve as a single-value metric summary of the $DSE(u, v)$.

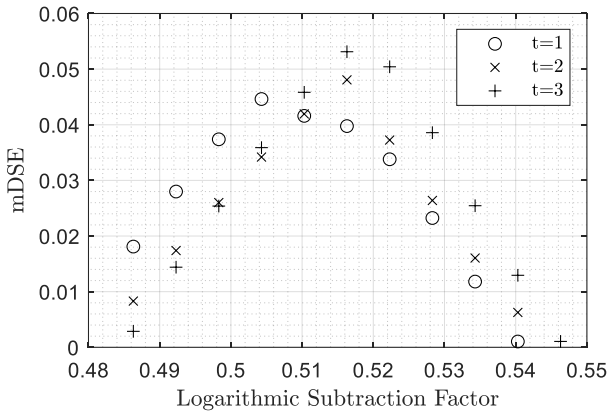

Fig. 2: Example system optimization using $mDSE$ as a quality metric. The soft-tissue subtraction factor was optimized for the system described in Table 3 for varying phantom $t$ parameter values. This example showcases the potential role of the $mDSE$ when designing dual-energy systems and image algorithms.

While the $DSE$ is able to describe the image properties mentioned above, one notable exception that this method is lacking is the presence and severity of motion artifacts. These are artifacts existing in multi-shot dual-energy systems that are caused by misalignments between the primary images, typically caused by patient motion between the acquisitions. Is it possible to ensure that the $DSE$ describes these edge artifacts (likely as a mid-frequency dip) by simulating patient motion through the moment of the phantom between the different primary image acquisitions or through the warping of one of the primary acquisitions by a known motion kernel. The details of such extensions to the $DSE$ are left as the subject of future work.

## VI. CONCLUSIONS

The proposed metric and its associated procedure allow for a quantitative and objective evaluation of the transfer of image quality properties between a conventional radiography image and a dual-energy image generated by the same system. We conclude that, alongside quality metrics dedicated to conventional radiography such as detective quantum efficiency, it is now possible to extend objective X-ray system evaluation to include dual-energy subtraction radiography.

We see the value of the *DSE* lying not only in its ability to characterize and compare existing dual-energy systems, but also in the assistance in can provide to the development of new systems. This is of particular importance due to the wide variety of existing techniques used to perform and augment the dual-energy subtraction process. Given the competing nature of the various image quality properties, it is easy to optimize for one of them while neglecting the rest. The *DSE* provides a picture which simultaneously addressed many of these properties, allowing for a more careful optimization of systems and algorithms.

REFERENCES

1. Vock P, Szucs-Farkas Z. Dual energy subtraction: Principles and clinical applications. *Eur J Radiol*. 2009;72(2):231-237. doi:10.1016/j.ejrad.2009.03.046

2. MacMahon H, Li F, Engelmann R, Roberts R, Armato S. Dual energy subtraction and temporal subtraction chest radiography. *J Thorac Imaging*. 2008;23(2):77-85. doi:10.1097/RTI.0b013e318173dd38

3. Jerrold T. Bushberg, J. Anthony Seibert, Edwin M. Leidholdt JMB. *The Essential Physics of Medical Imaging*. 3rd ed. Philadelphia, PA: LIPPINCOTT WILLIAMS & WILKINS; 2012.

4. Brody W, Cassel D, Sommer F, et al. Dual-energy projection radiography: initial clinical experience. *Am J Roentgenol*. 1981;137(2):201-205. doi:10.2214/ajr.137.2.201

5. Fischbach F, Freund T, Röttgen R, Engert U, Felix R, Ricke J. Dual-Energy Chest Radiography with a Flat-Panel Digital Detector: Revealing Calcified Chest Abnormalities. *Am J Roentgenol*. 2003;181(6):1519-1524. doi:10.2214/ajr.181.6.1811519

6. Gilkeson RC, Novak RD, Sachs P. Digital Radiography with Dual-Energy Subtraction: Improved Evaluation of Cardiac Calcification. *Am J Roentgenol*. 2004;183(5):1233-1238. doi:10.2214/ajr.183.5.1831233

7. Uemura M, Miyagawa M, Yasuhara Y, et al. Clinical evaluation of pulmonary nodules with dual-exposure dual-energy subtraction chest radiography. *Radiat Med*. 2005;23(6):391-397. http://www.ncbi.nlm.nih.gov/pubmed/16389980.

8. Szucs-Farkas Z, Patak MA, Yuksel-Hatz S, Ruder T, Vock P. Single-exposure dual-energy subtraction chest radiography: Detection of pulmonary nodules and masses in clinical practice. *Eur Radiol*. 2008;18(1):24-31. doi:10.1007/s00330-007-0758-z

9. Ide K, Mogami H, Murakami T, Yasuhara Y, Miyagawa M, Mochizuki T. Detection of lung cancer using single-exposure dual-energy subtraction chest radiography. *Radiat Med*. 2007;25(5):195-201. doi:10.1007/s11604-007-0123-9

10. Sabol JM, Avinash GB, Nicolas F, Claus BEH, Zhao J, Dobbins III JT. The Development and Characterization of a Dual-Energy Subtraction Imaging System for Chest Radiography Based on CsI:Tl Amorphous Silicon Flat-Panel Technology. *Med Imaging 2001 Phys Med Imaging*. 2001;4320:399-408. doi:10.1117/12.430897

11. Kashani H, Gang JG, Shkumat NA, et al. Development of a High-performance Dual-energy Chest Imaging System. Initial Investigation of Diagnostic Performance. *Acad Radiol*. 2009;16(4):464-476. doi:10.1016/j.acra.2008.09.016

12. McAdams HP, Samei E, Dobbins J, Tourassi GD, Ravin CE. Recent advances in chest radiography. *Radiology*. 2006;241(3):663-683. doi:10.1148/radiol.2413051535

13. Ishigaki T, Sakuma S, Horikawa Y, Ikeda M, Yamaguchi H. One-shot dual-energy subtraction imaging. *Radiology*. 1986;161(1):271-273. doi:10.1148/radiology.161.1.3532182

14. Ho J-T, Kruger RA, Sorenson JA. Comparison of dual and single exposure techniques in dual-energy chest radiography. *Med Phys*. 1989;16(2):202-208. doi:10.1118/1.596372

15. Lopez Maurino S, Badano A, Cunningham IA, Karim KS. Theoretical and Monte Carlo optimization of a stacked three-layer flat-panel x-ray imager for applications in multi-spectral diagnostic medical imaging. *Proc SPIE 9783, Med Imaging 2016 Phys Med Imaging*. 2016;9783:97833Z. doi:10.1117/12.2217085

16. Alvarez RE, MacOvski A. Energy-selective reconstructions in X-ray computerised tomography. *Phys Med Biol*. 1976;21(5):733-744. doi:10.1088/0031-9155/21/5/002

17. Cardinal HN, Fenster A. An accurate method for direct dual-energy calibration and decomposition. *Med Phys*. 1990;17(3):327-341. doi:10.1118/1.596512

18. Chuang K-S, Huang HK. Comparison of four dual energy image decomposition methods. *Phys Med Biol*.

1988;33(4):455-466. doi:10.1088/0031-9155/33/4/005

19. Lehmann LA, Alvarez RE, Macovski A, et al. Generalized image combinations in dual KVP digital radiography. *Med Phys*. 1981;8(5):659-667. doi:10.1118/1.595025

20. Sabol JM, Avinash GB. Novel method for automated determination of the cancellation parameter in dual-energy imaging: evaluation using anthropomorphic phantom images. In: Yaffe MJ, Antonuk LE, eds. Vol 5030. ; 2003:885. doi:10.1117/12.480195

21. Hinshaw DA, Dobbins III JT. Recent progress in noise reduction and scatter correction in dual-energy imaging. In: Van Metter RL, Beutel J, eds. *Medical Imaging 1995: Physics of Medical Imaging*. Vol 2432. ; 1995:134-142. doi:10.1117/12.208330

22. Richard S, Siewerdsen JH. Cascaded systems analysis of noise reduction algorithms in dual-energy imaging. *Med Phys*. 2008;35(2):586-601. doi:10.1118/1.2826556

23. Allec N, Abbaszadeh S, Scott CC, Karim KS, Lewin JM. Evaluating noise reduction techniques while considering anatomical noise in dual-energy contrast-enhanced mammography. *Med Phys*. 2013;40(5):3-6. doi:10.1118/1.4799841

24. Cardinal HN, Fenster A. Analytic Approximation of the Log-Signal and Log-Variance Functions of X-Ray Imaging Systems, with Application to Dual-Energy Imaging. *Med Phys*. 1991;18(5):867-879. doi:10.1118/1.596744

25. Alvarez RE. Dimensionality and noise in energy selective x-ray imaging. *Med Phys*. 2013;40(11):1-13. doi:10.1118/1.4824057

26. Alvarez RE, Seibert JA, Thompson SK. Comparison of dual energy detector system performance. *Med Phys*. 2004;31(3):556-565. doi:10.1118/1.1645679

27. Johns PC, Yaffe MJ. Theoretical optimization of dual-energy x-ray imaging with application to mammography. *Med Phys*. 1985;12(3):289-296. doi:10.1118/1.595766

28. Tanguay J, Kim HK, Cunningham IA. A theoretical comparison of x-ray angiographic image quality using energy-dependent and conventional subtraction methods. *Med Phys*. 2012;39(1):132-142. doi:10.1118/1.3658728

29. Richard S, Siewerdsen JH, Jaffray DA, Moseley DJ, Bakhtiar B. Generalized DQE analysis of radiographic and dual-energy imaging using flat-panel detectors. *Med Phys*. 2005;32(5):1397-1413. doi:10.1118/1.1901203

30. Richard S, Siewerdsen JH. Comparison of model and human observer performance for detection and discrimination tasks using dual-energy x-ray images. *Med Phys*. 2008;35(11):5043-5053. doi:10.1118/1.2988161